\documentclass[10pt,preprint]{aastex}

\usepackage[cmex10]{amsmath}

\usepackage{sansmathfonts}
\usepackage[T1]{fontenc}
\usepackage{enumitem}
\usepackage{graphicx}
\usepackage{hyperref}
\hypersetup{colorlinks=true,linkcolor=blue,citecolor=blue,filecolor=blue,urlcolor=blue}

\newcommand{\ms}{\mbox{m s$^{-1}$}}

\begin{document}

\title{A small actively-controlled high-resolution spectrograph based on off-the-shelf components}

\author{H.R.A. Jones\altaffilmark{1}, W. E. Martin\altaffilmark{1}, G. Anglada-Escudé\altaffilmark{2}, R. Errmann\altaffilmark{1}, D. A. Campbell\altaffilmark{1}, C. Baker\altaffilmark{3}, C. Boonsri\altaffilmark{4}, P. Choochalerm\altaffilmark{1}
\altaffiltext{1}{Centre for Astrophysics Research, University of Hertfordshire, College Lane, Hatfield AL10 9AB, UK}
\altaffiltext{2}{Institute of Space Sciences, Campus UAB, Carrer de Can Magrans, s/n 08193 Barcelona, Spain}
\altaffiltext{3}{Queen Mary University of London, Mile End Road, London AL10 9AB, UK}
\altaffiltext{4}{National Astronomical Research Institute of Thailand, 260 Moo 4, T. Donkaew, A. Maerim, Chiang Mai, 50180, Thailand}
}

\begin{abstract}
We present the design and testing of a prototype in-plane echelle spectrograph based on an actively controlled fibre-fed double-pass design. This system aims to be small and efficient with the minimum number of optical surfaces - currently a collimator/camera lens, cross-dispersing prism, grating and a reflector to send light to the detector. It is built from catalogue optical components and has dimensions of approximately 20$\times$30 cm. It works in the optical regime with a resolution of $>$70,000. The spectrograph is fed by a bifurcated fibre with one fibre to a telescope and the other used to provide simultaneous Thorium Argon light illumination for wavelength calibration. The positions of the arc lines on the detector are processed in real time and commercial auto-guiding software is used to treat the positions of the arc lines as guide stars. The guiding software sends any required adjustments to mechanical piezo-electric actuators which move the mirror sending light to the camera removing any drift in the position of the arc lines. The current configuration using an sCMOS detector provides a precision of 3.5 milli-pixels equivalent to 4~m/s in a standard laboratory environment.
\end{abstract}


\section{Introduction}

The achievement of meter-per-second radial velocity precision is one of the major technological breakthroughs of recent decades. Although this effort is sometimes viewed as driven by the search for extra-solar planets following the discovery of the first promising signals (e.g., Latham et al. 1989, Wolszczan \& Frail 1992), the quest for highly accurate radial velocity measurements had been under development for many decades. The path from the suggestion that stellar velocities might be accurately calibrated (Struve 1952), to solar studies (Becker 1976; Koch \& Woehl 1984) to pioneering stellar observations using Hydrogen Fluoride gas cells (Campbell \& Walker 1979; Campbell, Walker \& Yang 1988) to sub-\ms with ESPRESSO (e.g., Pepe et al. 2020), long-term several \ms ~RMS (and short-term sub-\ms ~measurements with ESO~3.6m/HARPS (e.g., Pepe et al. 2000), Keck/HIRES, e.g., Vogt et al. 1994, Butler et al, 1996),  Magellan/PFS (e.g. Crane et al. 2006), HET/HRS(e.g., Tull 1998), TNG/HARPS-N (Sosnowska et al. 2012) and VLT-UVES (e.g., Butler et al. 2019) and is one that leads through several generations of similar spectrographs, detectors, and calibration methodologies.  

Although the instruments have not changed dramatically, the ongoing improvements in the production of accurate and precise radial velocities has arisen through the combination of a number of factors. That is, keeping instruments stable over long periods of time along with the ability to accurately calibrate spectra and to effectively reduce astronomical spectra. However, the large beam sizes of large telescopes has tended to mean that large bespoke optics are required for the optical elements, e.g., HIRES on Keck uses a mosaic of three gratings with a size of 1$\times$4 foot (Vogt 1994). The need for mosaic gratings and bespoke large optics continues to contribute to the ongoing high cost of such instruments, e.g., 23.7~MEuros\footnote{\url{http://obswww.unige.ch/~pepe/Repository/DRD03\_Executive\%20Summary\_VLT-TRE-ESP-13520-0037\_Iss3.pdf}} for ESPRESSO and might further escalate as such instruments are built for larger telescopes.

We contribute to this series of work by investigating some alternative approaches for the configuration of a high-resolution spectrograph.  We describe our experiments with a laboratory-based, test-bed, high-resolution spectrograph using traditional components. 
The goal of this paper is to demonstrate new design features which lead to a reduction in the size and cost without modest impact on performance. In the long term it can be envisaged the popularity of high resolution spectrographs will continue to build particularly if their cost/performance can be improved so that they can be deployed much more widely and following the concept of Lovis et al. (2017) offer one route to exploit the potential for high resolution integral field spectroscopy behind a high resolution adaptive optics instrument. 
 
Some of our new design features are to be deployed and built into the as EXOhSPEC (EXOplanet High resolution SPECtrograph) instrument for the 2.4m Thai National Telescope. The spectrograph is envisaged as fiber-fed with a resolution of greater than 70,000 and designed to be a workhorse high resolution spectrograph. The optical design and Zemax tolerancing analysis of this spectrograph is described in Lhospice et al. (2019) and the finite element analysis for the collimator design in Kawinkij et al. (2019). Our currently preferred methodology for achieving efficient focal reduction is through efficient tapering of the input fibre from the telescope as presented by Choochalerm et al. (2020). Alongside these efforts we have developed a data reduction pipeline (Errmann et al. 2020) which can be used for EXOhSPEC but is specifically built with the flexibility to examine data from our laboratory prototypes. In Section 2, we outline and present the basic spectrograph, considering its design and components. In Section 3, we describe our initial set of components and first test results considering modal noise, temperature, throughput and active stabilisation.

\section{Outline design}
\begin{figure}
\begin{center}
\includegraphics[height=100mm,angle=-0]{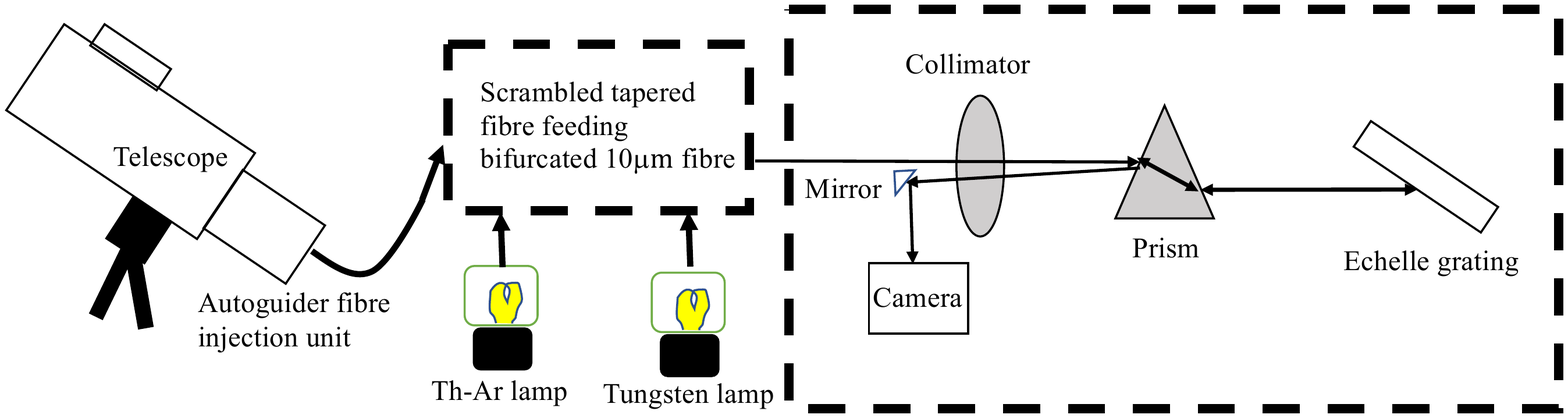}\vspace{-2cm}
\includegraphics[width=120mm,angle=-0]{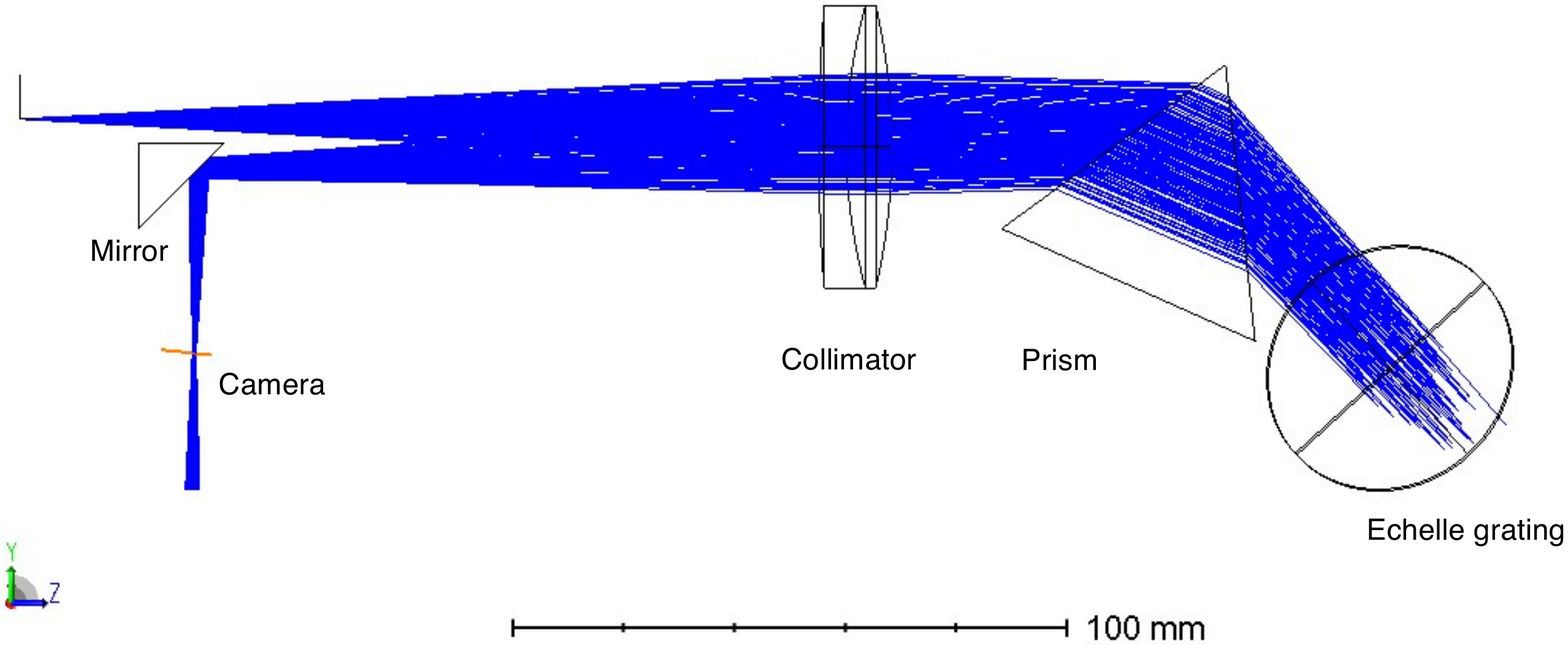}
\caption{{\bf Top -- schematic of the components of the instrument. A variety of different options are available within the fibre feed arrangements to the spectrograph. Once the fibre enters the spectrograph the light path is fixed with double passes of the collimator and the prism. The angular changes within collimator and dispersing prism are exaggerated for illustration. Bottom -- optical layout from a Zemax model is illustrated. The figure is a top view in the plane of the echelle cross-dispersion plane. The black outlines are schematics of the optical components. }} 
\label{zemax}
\end{center}
\end{figure}

\begin{figure}
\begin{center}
\includegraphics[height=80mm,angle=-180]{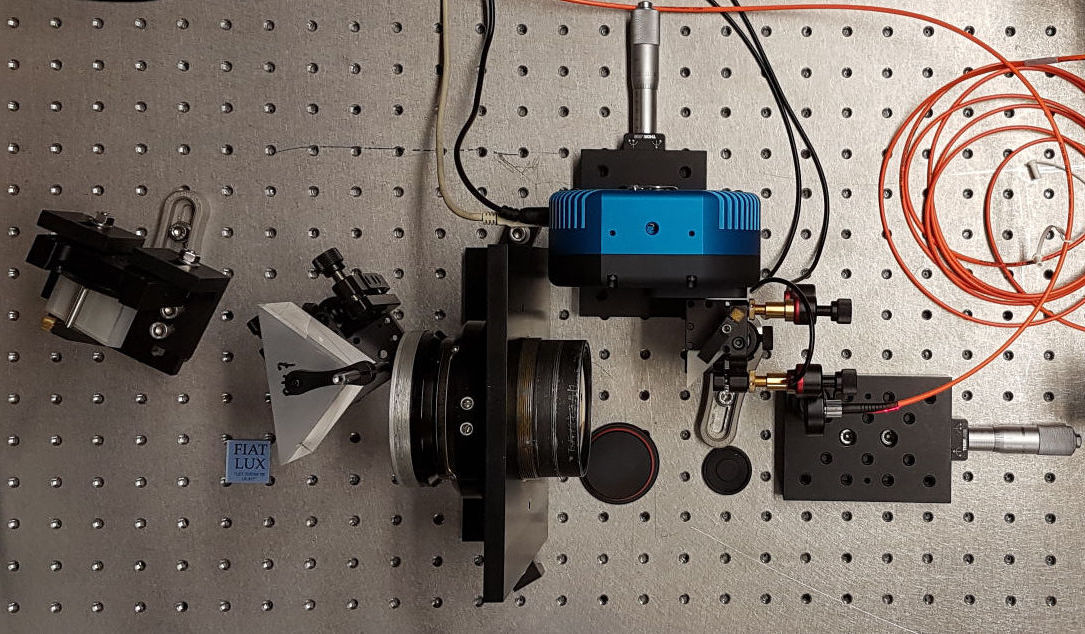}\\
\includegraphics[height=30mm,angle=-0]{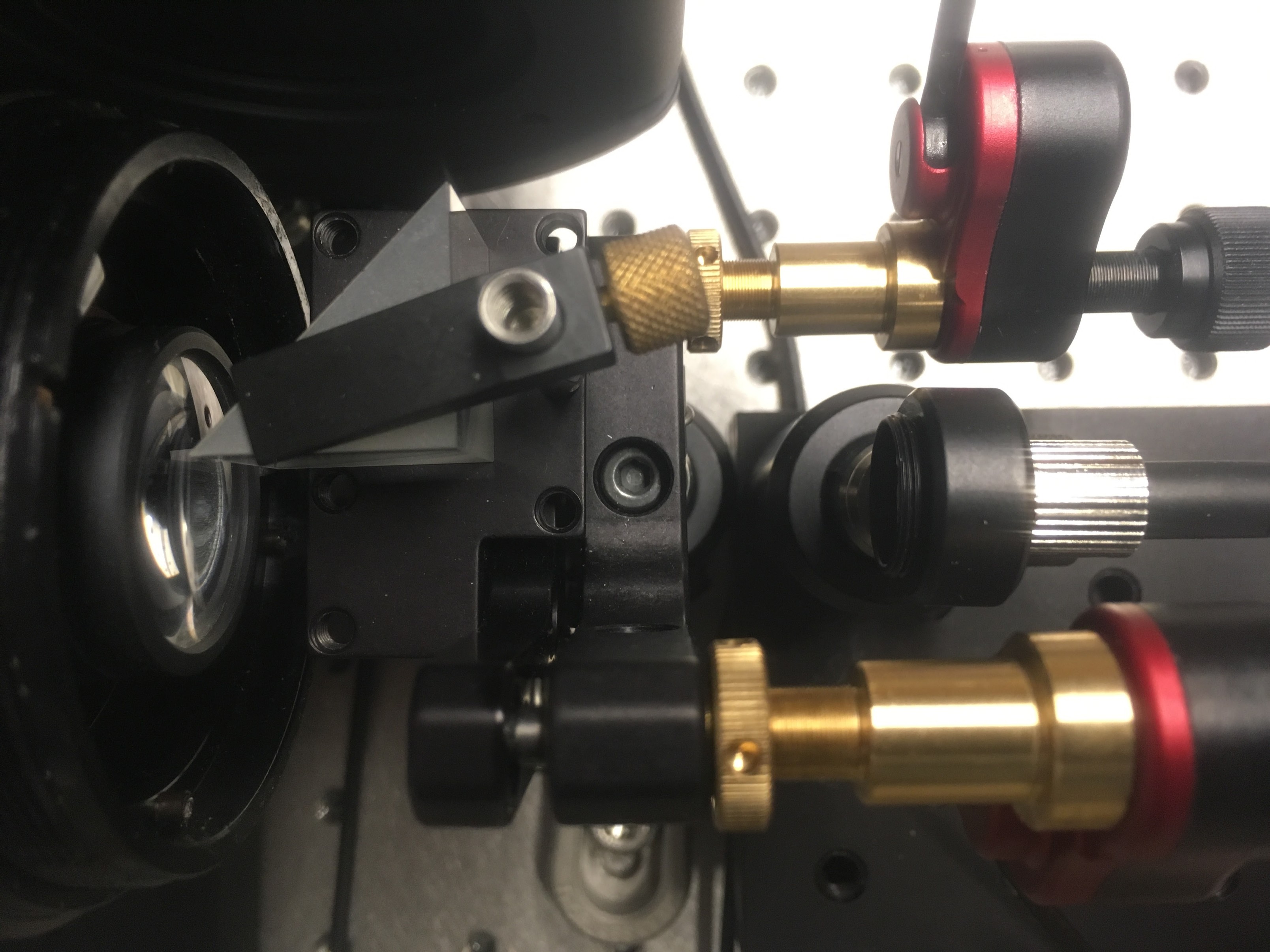}
\includegraphics[height=30mm,angle=-0]{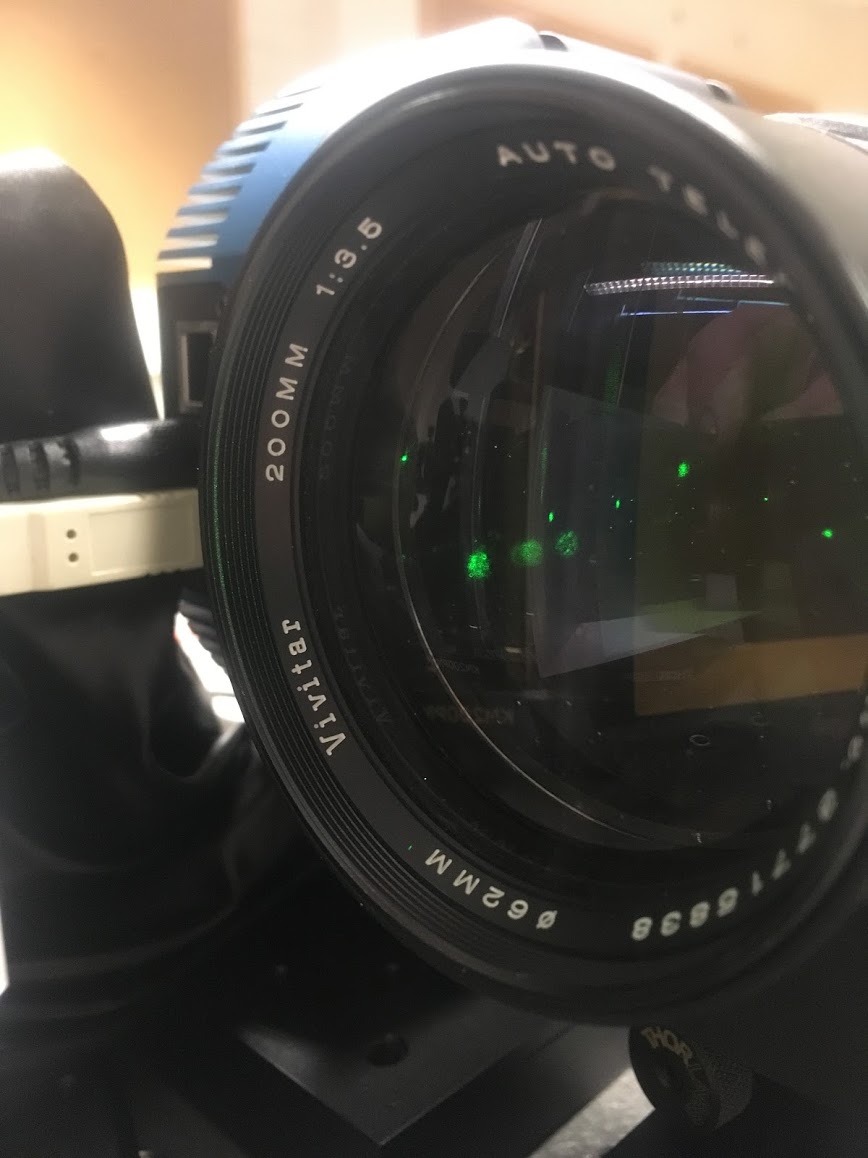}
\includegraphics[height=30mm,angle=-0]{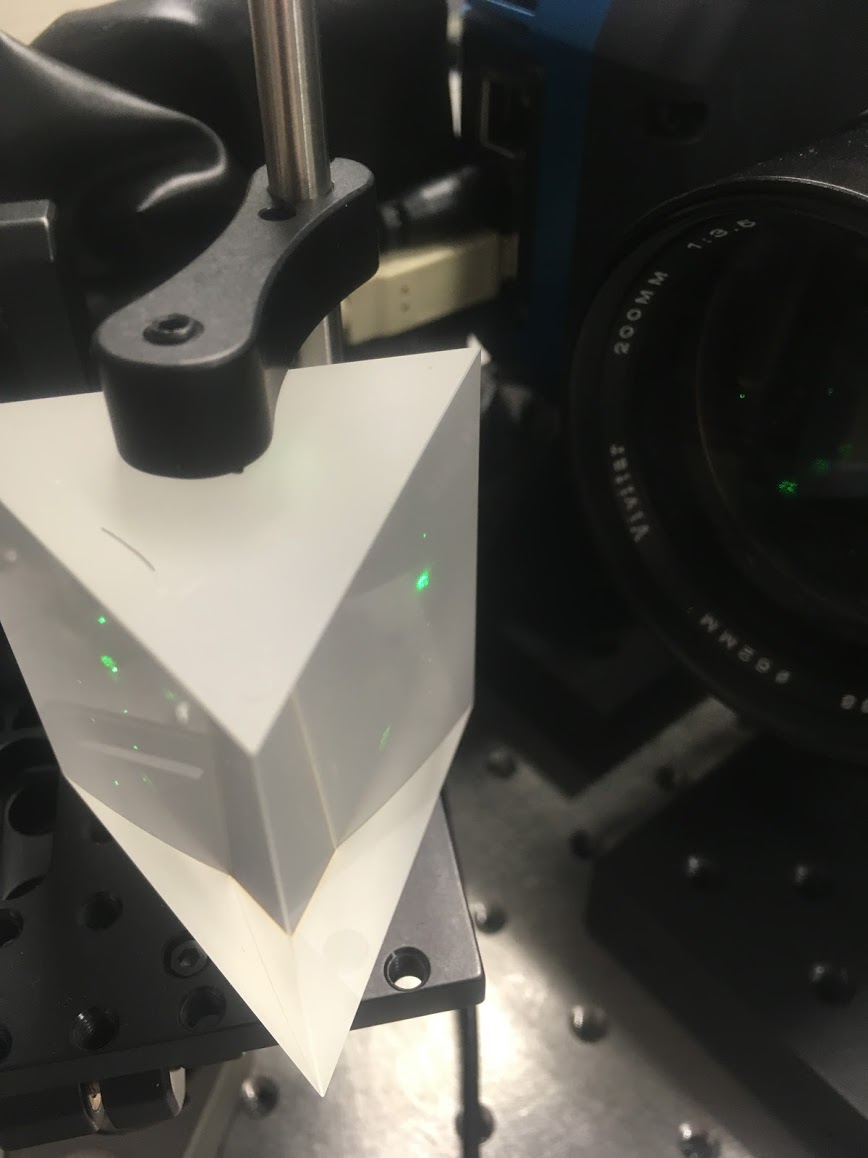}
\includegraphics[height=30mm,angle=-0]{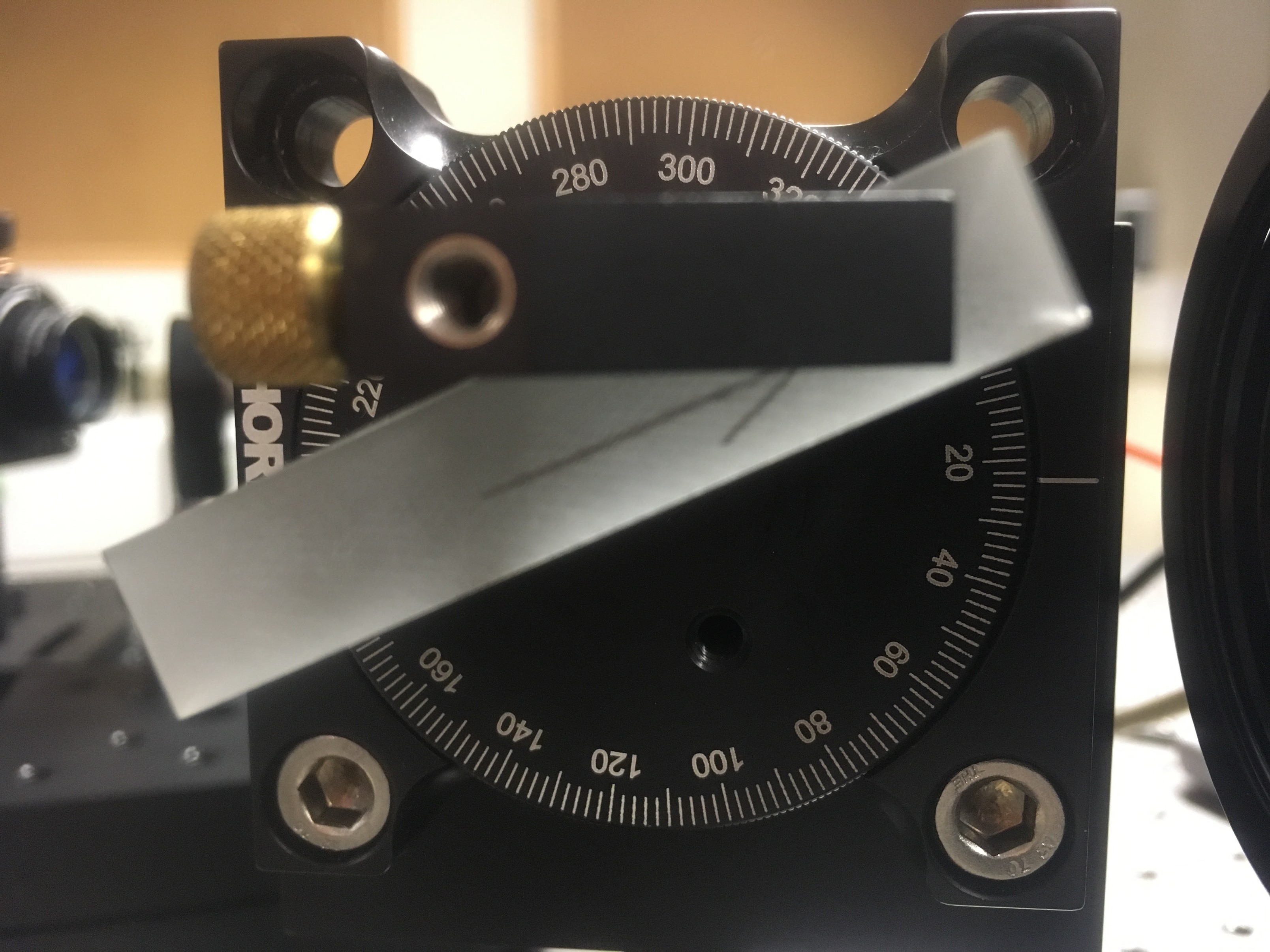}
\includegraphics[height=30mm,angle=-0]{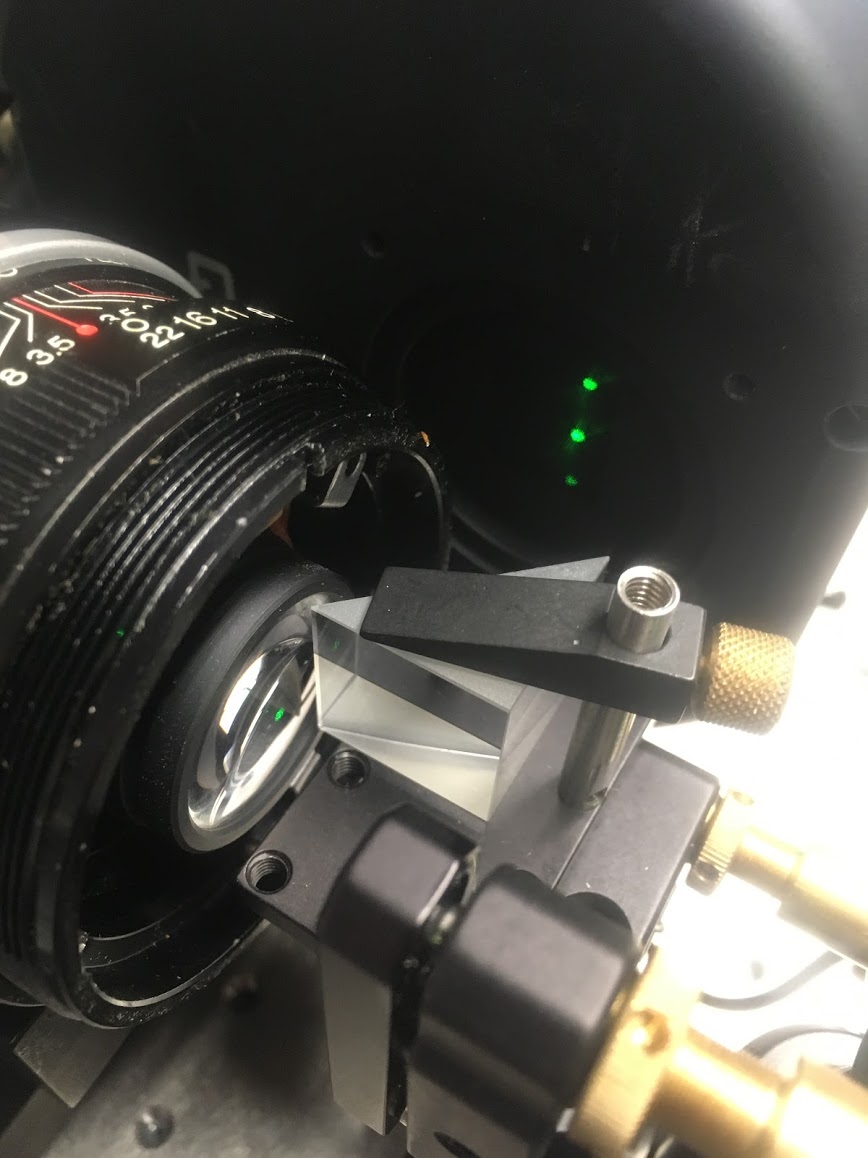}
\caption{The image shows the layout of instrument on the bench. Following the light path: the fibre input (orange) is towards the top-left, it passes through the camera lens top-middle, the cross dispersing prism top-right, grating middle-left. The light then returns by almost the same path and is reflected by the pickoff prism mirror immediately before the camera in the bottom-left. The lower images show the individual optical components from left to right: fibre input (on the right-hand side with red, gold and black piezo actuators above and below and also showing camera lens on left-hand side), camera lens - Vivitar, 200mm f3.5 telephoto lens, cross-dispersing prism (and camera lens), grating, pickoff mirror (a silvered prism, with camera lens to the left and detector to the right illuminated with laser diode diffraction spots).} 
\label{layout}
\end{center}
\end{figure}

The construction of a high resolution spectrograph aimed at measuring precision radial velocities on a very limited budget, required that our pathfinder spectrographs work in the optical regime where detectors, optical equipment and their coatings are much more affordable. 
The spectrograph uses standard optical mounting hardware on an \hbox{$4 \times 2$ foot} optical bench. It is located in the Applied Optics Laboratory on the College Lane campus of the University of Hertfordshire. In Fig. $\ref{zemax}$ we illustrate a schematic layout and the outline Zemax design of the spectrograph. In Fig. $\ref{layout}$ we illustrate one of our initial implementations of the spectrograph along with the various components of the spectrograph.

\subsection{Telescope and laboratory}
We have arranged the laboratory to be fed by fibre optic cable from a relatively sheltered nearby first floor flat roof of an adjacent building. There we added a Celestron 8~inch telescope, with a SkyWatcher EQ3-2 Pro Synscan mount. Although it is in a highly light-polluted area of campus, overlooked on all sides by office windows, suffers from vibrations caused by a nearby heat exchanger and with a rather restricted view of the sky it can in principle provide access to the Sun, Moon and potentially bright stars. We attach a core graded-index 50~$\mu$m FC/PC connectorised fibre to the telescope using a Shelyak fibre injection unit (PF0018 F/6 50$\mu$m injection unit and M42 T adapter to the telescope). \\

The laboratory temperature control is tied to a central building control system and the ambient temperature of the room is constrained to within approximately 2$^\circ$C on a variety of timescales so we monitor temperature and humidity changes at the apparatus. In particular, the issue of hard to control atmospheric changes has encouraged us to investigate active control as a solution for the spectrograph stability. This can be thought of as an alternative solution to the usual deployment of precise high resolution spectrographs into a vacuum tank.

\subsection{Active stablisation}

Feedback loops are widely deployed in astronomy. These are widely found in telescope guiding systems where a variety of closed loop servo-actuated systems are used to optimise the response of a telescope to the wind and other environment noise. This allows them to maintain precise pointing position on the sky. Such techniques are also sometimes used in optical astronomy. For example, the `ultra-stable' cryostat used by the ESPRESSO instrument, e.g., Lizon et al. (2016) provides continuous optical centroiding corrections based on temperature and pressure sensors inside the detector dewar. This enables the detection and correction of instability and drift of image motion. This detector system provides 2nm RMS centroid stability and is a key part of the long-term 50 cm/s precision achieved by ESPRESSO (e.g., Pepe et al. 2020). Ultimately, it might be to deploy some of the techniques developed by the atom interferometer experiments (e.g., Peters, Chung \& Chu 2001) which have led to a measured noise of better than 10$^{-19}$~m in differential arm length over a wide range of frequencies (30--3000~Hz) by the gravitational wave experiments (e.g., Buikema et al. 2020) providing a particularly stark example of the power of the application of closed loop technology (e.g., Bond et al. 2016). 

Conventional designs of high-precision echelle spectrographs rely principally on good passive athermal designs and stable mechanical components.  Environmental variables such as temperature, pressure and vibration are isolated from the spectrograph by insulation, inert gas filling or vacuum operation and vibration damping arising from massive and/or highly damped materials. The addition of active control of a few parameters such as distances between components or deflection angles of certain components can relieve the need for expensive and/or massive components to achieve the same level of overall stability.  An ‘active metrology’ scheme whereby, for example, the distance between two components is measured with (commercially available) laser interferometers and used to drive a piezoelectric motion transducer to stabilise the optical distance can offer compensation for both temperature and pressure effects in the spectrometer.  With careful passive design and optimum optical configurations, this approach might offer considerable benefit and potentially offer a retrofit solution to improve the performance of existing spectrometers.

A complementary approach to athermal design and stable mechanical components is to wonder how to produce a stable spectrum position on the sensor? If a reference spectral source is imaged through the spectrometer then the position of this (these) source(s) in the dispersed image on the detector can be used to produce an error signal to drive a component of the spectrometer so that the image remains fixed in position on the detector.  If the object spectrum is imaged through the same optical path then it too is stabilised. Note that to zeroth order it does not matter what is between the spectrometer input and output.  Obviously it will be good design practice to have a stable optical/mechanical design in order to reduce the range of the control corrections needed. To first order, the spectrum on the detector is stabilised to the limiting stability of the reference source.  The remaining issues for stability will be second and third order effects due to two principal factors: a) the angle/position of the optical path through the spectrometer will be changing with the corrections and b) changes in the dispersion of the spectrometer.  There will be slight differences in the design approach to minimise aberrations from third order effects such as the change of dispersion of the lens and prism with temperature and of the surrounding gas environment.  Changes in the dispersion of the spectrometer may be approached in a similar way that achromatic lenses are designed, i.e. introduction of components which cancel the overall dispersion variations with temperature or pressure. The two primary requirements for this design are a usable stable reference source and a control loop with sufficient capability to affect the necessary corrections.  So for example, ThAr lamp lines which are routinely used as the reference sources for precision spectrographs can be used as the reference source for such active correction and this is a methodology we explore.

\subsection{Preliminary optical design}


One of the key issues for high precision spectrographs is that they must be stable on all timescales which might be of interest for astronomical phenomena. When considering exoplanets this might be from hours to hundreds of years (e.g., Feng et al. 2019). Both simultaneously calibrated spectrographs such as HARPS and Iodine stablised spectrographs such as HIRES work on the premise that the optical arrangement should be as stable as possible. So in the case of common user spectrographs like HIRES, a bespoke calibration procedure is required to ensure that each night begins and ends with ThAr lines falling on exactly the same position on the detector. HARPS achieves this with extreme temperature and pressure control along with complete isolation of the spectrograph from any human interaction.

There are many different successful designs available for us to develop from. We seek to benefit from these and in particular we take inspiration from the HARPS (Pepe et al. 2000) and PFS (Crane et al. 2006) instruments due to their exquisite performance over more than a decade of operation. Both of these designs are themselves derived from several similar previous instruments with the fundamental designs and grating and echelle relationships being laid out in the literature, e.g., Schroeder (1967), (1970), Schroeder \& Anderson (1971). Given that we have a laboratory available some distance from any telescope, a fibre feed like HARPS was desirable. On the other hand, the small size and re-use of optical components provided by a double-pass design was particularly appealing from the PFS design.

\begin{table}[t]
{\small
\caption{Beam diameter, grating length and ruled area requirements for 
different fibre and grating parameters. The first line of the table is given in italic and approximately reproduces the effective size of the HARPS beam based on its 84$\times$21.4cm grating used at a blaze angle of 75 deg mosaic. A blaze angles of 63 degrees is also chosen as a representative case, e.g., Thorlabs GE-2550-0363.
\label{tab:examples}}
\begin{tabular}{cccc|ccc}
\\
 Resolution & Fibre    & Numerical & Blaze      & Beam       & Grating & Ruled \\ 
            & diameter & aperture & angle      & diameter   & length  & Area  \\ 
\hline
 $[\lambda/\delta \lambda]$      & [$\mu$m] & $[-]$     & [deg]      & [cm]       & [cm]    & [cm$^2$] \\
\hline
 ${\it10^5}$     & $\it {50}$       &  $\it {0.22}$     & $\it {75}$         & $\it {19.7}$       & $\it {76.0}$    & $\it {1498}$   \\
 $10^5$     & 10       &  0.22     & 75         & 4.0        & 15.2    &  60.8  \\
\hline
$10^5$     & 50       &  0.22     & 63         & 37.5       & 82.4   &  3088  \\
 $10^5$     & 10       &  0.22     & 63         & 7.6       & 16.5   &  125  \\
\end{tabular}
}
\end{table}

One of the major cost items of a high resolution spectrograph is usually its echelle grating. In the case of HARPS, its grating is constructed from two gratings mosaiced together to give a total area of 84$\times$21.4 cm. In order to scale down the required grating size  we can make use of the well established literature results that the resolution of a spectrograph is proportional to the angle of the slit opening (fibre diameter) as seen from the collimator (e.g., Schroeder 1987). Given that we are focussed on a high resolution instrument, we choose a resolution of 10$^5$ close to that of HARPS in Table 1 to show a HARPS-like configuration as well as the case for a catalogue Thorlabs grating (GE-2550-0363) to emphasize that by making the fibre diameter smaller for a fixed numerical aperture and blaze angle we can use a smaller echelle grating\footnote{Presented in more detail at \url{http://star.herts.ac.uk/exohspec/design_outline.pdf}}.

Thus taking the HARPS spectrograph as a model we might hope to emulate its resolution and approximate optical characteristics based on line 2 or 4 of the table by ensuring that the numerical aperture of the beam is preserved say when moving from a 3.6m telescope to a 0.7m telescope by rescaling the size of the input beam accordingly by a factor of five (3.6/0.7). Geometrical optics suggests that this would lead to a loss of light proportional to the change in area of the beam size (e.g. Haynes et al. 2012) resulting in most of the light being lost. However, based on the work of Choochalerm et al. (2020) we might use suitably tapered graded index fibres to reduce this beam size by a factor of 5 with perhaps less than a factor of two efficiency loss.





\section{Spectrograph components}
We now discuss our consideration for different spectrograph components used for our tests along with their approximate transmissions in Table 2. A schematic of the apparatus is given in Fig. \ref{zemax}.

\subsection{Echelle grating}

Following Section 2.2, with the reduction of size of the required grating by a factor of five fits approximately with available off-the-shelf gratings like the Newport 53019ZD01-413E (27~grooves/mm, 70$\deg$ blaze) with dimensions: 6x15cm for a few thousand dollars or with reduced resolution using the volume produced Thorlabs GE-2550-0363 with 31.6~grooves/mm, 63$\deg$ blaze, 2.5$\times$5.0~cm for a few hundred dollars. The echelle is mounted facing downwards in order to minimise exposure to dust.

This brings a consequent reduction in the size of the lens and cross-disperser optics to the scale of standard catalogue components and in particular the utility of standard telescope lenses. While this has the potential to yield very significant cost savings and clarity of performance due to using well-characterised mass produced components, there are a number of issues that might arise from doing this. For example, the camera optics have to be of high quality in order to keep the aberrations smaller than the fibre image size. In order to investigate these and other potential other issues we have built several prototypes.

\begin{table}[t]
{\footnotesize
\caption{The approximate performance of main spectrograph components noting that most have very considerable wavelength dependence and performance trade-offs. The  acquisition and guidance of light from the telescope is omitted from this table.}
\begin{tabular}{c|cc}
\\
Component & Manufacturer Catalogue Number & Transmission (\%)   \\ 
\hline
Fibre feed & FG010LDAFBUNDLE & $$75 \\
Scrambling & Tektronix AFG1022 & $$50 \\
Echelle Grating & Newport 53019ZD01- 413E (27~grooves/mm, 70$\deg$ blaze) & 40 \\
Collimator  & Thorlabs AC254-050-B and AC254-200-B & $$90 \\
Cross Dispersing Prism  & 80mm N-BK7 60$\deg$ (modified Thorlabs PS854)& $$50 \\
Turning mirror & Thorlabs MRA20-G01 & $$95 \\
Camera & Photometrics Prime BSI sCMOS (2048$\times2048$, 6.5 $\mu$m pixels)& $$70 \\
\hline
Overall  & 4370-7400\AA, Resolution 82500, Sampling 4--13 pix per FWHM & $$4 \\
\end{tabular}
}
\end{table}

\subsection{Collimator -- Camera Lens and Cross Dispersing Prism}
The spectrograph optics are used in double-pass and so act both as collimator and camera lens. The required optics are all spherical. Our setup has thus far been with off-the-shelf camera lenses (such as the 200mm Vitkar lens shown in Fig. \ref{layout}) or the achromatic doublet collimator and focussing lenses (from Thor Labs with 650-1050 nm coatings, AC254-050-B and AC254-200-B).  

In our double-pass configuration, the quality and performance of this lens is rather critical. The camera optics do have to be of high enough quality in order to keep the aberrations smaller than the fibre image size. Any significant misalignment or chromaticity is easily seen in the output spectra where the line shape of calibration lines can be compared between the centre and the edges. Our current camera layout does suffer from some coma and this variation of the spot shape over the camera field imposes a limits on the resolution and achievable radial velocity precision in our system. We have performed Zemax calculations and find that the coma we find is in line with the expectation for doublet lenses. Details of similar calculations for an improved triplet design are presented in detail in Lhospice et al. (2019) and Kawinkij et al. (2019). We also have a separate project to deliver a prototype based on a highly corrected small double Gauss lens which is laid out elsewhere, e.g., Baker et al. (2019). While the camera lens does limit the utility of our current setup, the current achromatic doublet implementation was judged sufficient for our initial tests.

We use an equilateral dispersive prism in order to cross disperse the beam. We have used a few different dispersing prisms particularly the Thorlabs PS854 (F2 50 mm). We now use a modified version of this, a N-BK7 anti-reflection coated 60$\deg$~prism (F2 80mm).
In principle a VPH grating (e.g., Seifahrt et al. 2018) would offer improved throughput but in the development stage prisms are readily available, easy to handle and offer a variety of dispersions, sizes and coatings.

\subsection{Detector}

A key driver in spectrograph design is the detector. The HARPS detector\footnote{\url{http://www.eso.org/sci/php/optdet/instruments/harps/FDR-Harps-detectors.pdf}} choice of two EEV type 44-82 CCDs with 15$\mu$m pixels leads to a pixel oversampling of around 4. These relatively large pixels and apparently excessive oversampling are a key element in the success of HARPS in the sense that it can utilise CCDs where every pixel produces high dynamical range and that the point-spread function is always fully sampled and far less prone to detector inhomogeneities and irregularities (e.g., Butler et al. 2019). In order to match these characteristics we need a large format camera with around 3$\mu$m pixels. Such a small size is rather smaller than supplied for conventional CCD detectors. In comparison with the HARPS CCDs, the newer E2V 290-99 in use for ESPRESSO, PEPSI, EXPRES and MAROON-X has approximately fivefold increase in pixels for an individual device, a reduction from 15 to 10 micron pixels for similar well depth — gain — read-out noise configurations, improved quantum efficiency by around 10\% (over the relevant wavelength range) with apparently similar linearity, charge transfer efficiency and cosmetic properties (Calderone et al. 2016). While these devices have excellent performance characteristics, the range of applications that they are used for is relatively limited and so professional astronomical grade CCDs sold by companies such as Teledyne (E2V) have a unit costs in excess of 100k dollars each. In recent years an alternative image recording technology CMOS has become widely available with a relatively low cost per pixel in comparison to CCDs. While scientific grade CMOS devices (sCMOS) have rapidly improved they have drawbacks for astronomical purposes. sCMOS can have high read noise and dark current, they generally have a rather small dynamic range and perhaps significant intrapixel non-uniformities (Zhan, Zhang \& Cao 2017). However, some of these sCMOS devices are deemed as mature enough to be used for future space missions such as JUICE (e.g. ESA 2014). 

\begin{figure}
\begin{center}
\includegraphics[width=150mm,angle=-0]{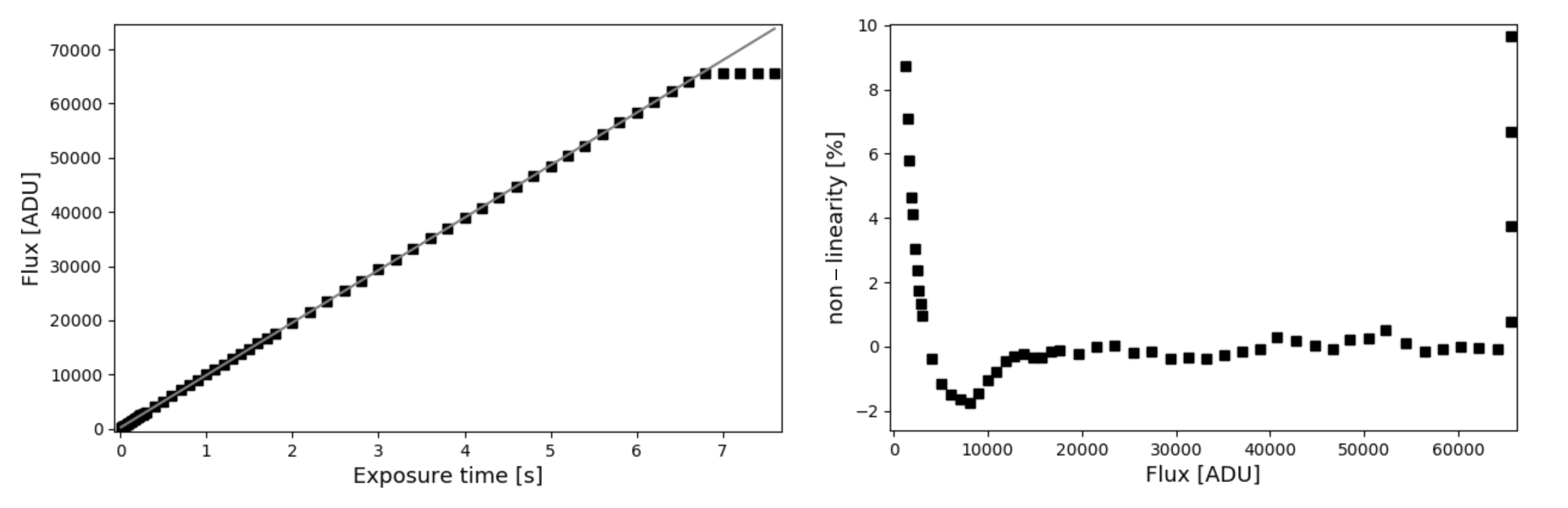}
\caption{Linearity of the Photometrics PrimeBSI camera shown in the left-hand plot for the full well depth and in the right-hand plot for the flux range between the two amplifiers. The flux was measured in the central 48x48 pixel sized area for different exposure times (black dots). For each exposure time, five flat field images were taken and then normalised and median combined. A linear fit was applied to the data points (grey line). The non-linearity shows the deviation of the data points to the linear fit, scaled by the flux.}
\label{cmos}
\end{center}
\end{figure}

For our experiments we have used (1) a QSI 532ws-M1 camera with 2184$\times$1472 6.8$\mu$m pixels and (2) a reburbished Photometrics Prime BSI sCMOS with 2048$\times$2048 6.5$\mu$m pixels. These have been set to typically give 70 orders with a wavelength coverage from 460--880 nm and a resolution 0.0015 – 0.0022~nm/pixel. Different order spacings and numbers of orders readily achieved by using different cross-dispersing prisms. The specifications of the CCD are somewhat better in most respects than the sCMOS but the negligible read time of the sCMOS makes it very convenient in a laboratory setting. However, the use of two amplifiers (for each pixel) to produce the 16-bit output of the sCMOS gives rise to significant non-linearity, plotted in Fig. \ref{cmos}. Photometrics (2020, private communication) have investigated different versions of their firmware which can improve the non-linearity to 2\%. Although this is linearity is comparable to the CCD, the CCD non-linearity is smooth whereas the sCMOS has a more complex correction of the functional form of the non-linearity between the amplifiers. A correction routine in the data processing code is used to linearise the signal. The cameras are both air or water cooled and are typically run at --30$^{\circ}$C from MaxIm DL version 6.18. Both cameras take in excess of an hour for their temperatures to stabilise.

\subsection{Spectrograph setup and alignment}
In order to align our spectrograph, we follow the following sequence of steps: (1) Use a green laser diode (e.g., Thorlabs CPS532 Laser diode module, 532~nm) to set up a straight path (constant height, aligned with the optical bench). (2) Insert the prism, rotate it and align so that the reflected/transmitted light doesn’t change height. (3) Insert the lens, adjust the height, position, and angles so that all surface reflections of the lens hit the laser. (4) Find the minimum deviation angle of the prism: rotate and observe the minimum spot deviation. (5) Insert the grating so that the spot hits approximately the centre of the grating and rotate it so that the reflected orders are approximately in the plane of the laser. (6) Insert the fibre holder, so that the laser passes through the center. (7) Attach the fibre illuminated with a red laser diode (e.g., Thorlabs CPS635S Laser diode module, 635 nm) and find the focal spot (the spectrum of the laser diode is in focus at the fibre end). (8) Move the fibre 2mm out of the centre. (9) Centre the grating and check that the reflection hits the reflection grating. (10) Centre the spectrum on the camera. (11) Put white light through the system to focus the camera. A rotation of the camera might be necessary. (12) The final focusing is done using the emission lines (ThAr), measuring the longitudinal position of the camera when either blue or red lines are in focus. The camera is then tilted accordingly. With the doublet lens, a tilt about 20 degrees is required.

\subsection{Data Reduction and Analysis}

The spectrograph data reduction system has been written for the flexible reduction of  cross-dispersed high-resolution data, such that its order tracing is as flexible as possible, in order to accommodate a variety of different components and setups. For example, the software is designed to cope with a variety of resolutions, order separations, overlaps and curvatures; as well as over/under sampling. It can be used with any type of calibration source once an initial wavelength solution has been found  through the identification of some known-wavelength spectral lines. The philosophy has been that once a wavelength solution exists for one configuration a new nearby solution might be found relatively easily given knowledge of the expected known wavelength spectral lines. The data reduction software known as HIFLEx is available on github\footnote{\url{https://github.com/ronnyerrmann/HiFLEx}} and described in Errmann et al. (2020). 

\subsection{Calibration}

\begin{figure}
\begin{center}
\includegraphics[width=150mm,angle=-0]{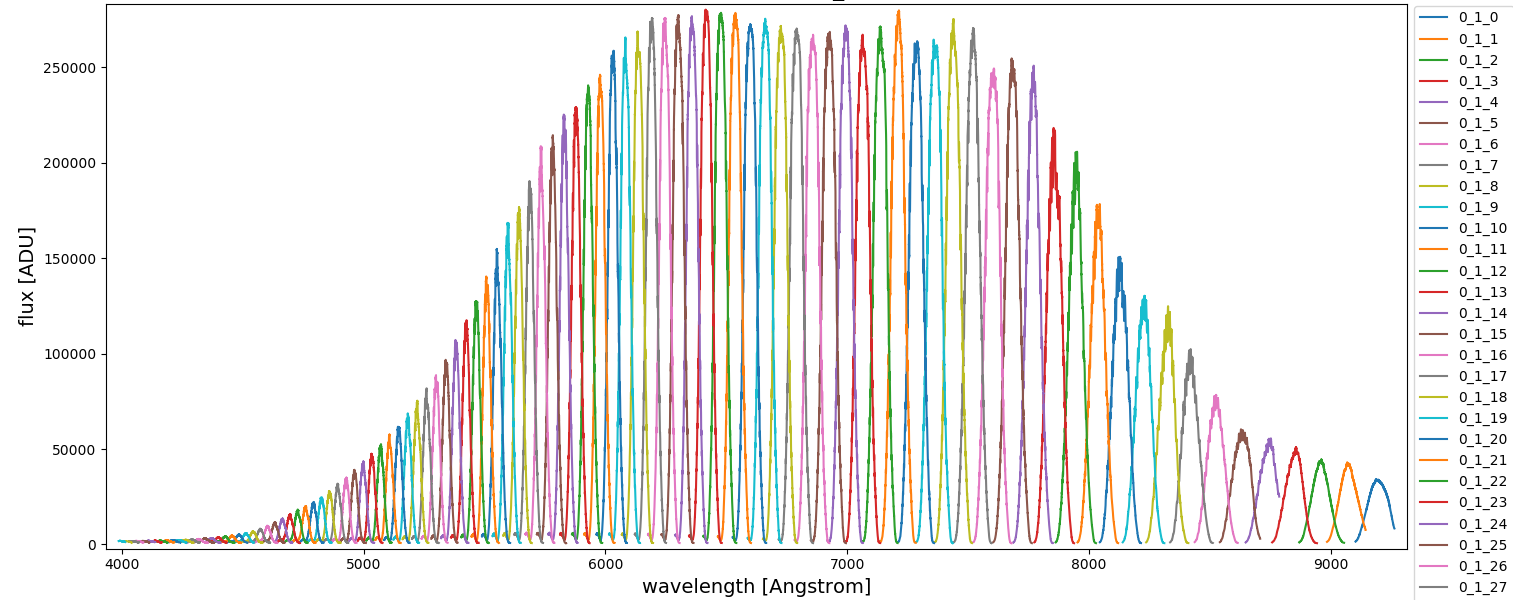}
\caption{The plot shows overlapping spectral orders producing using a flatfield lamp. }
\label{orders}
\end{center}
\end{figure}

For order tracing and flat fielding we use a stablised Tungsten lamp (Thor Labs SLS201L/M). Initially we used a so-called ``high intensity fiber light source'' (Thorlabs OSL1-EC), however, this lamp had rather limited red throughput. The relative system efficiency of the Tungsten lamp along with the camera and fibre throughput can be seen in Fig. $\ref{orders}$, which illustrates the overlapping spectral orders provided using the flatfield lamp. However, given the strong decline in efficiency towards blue and red spectral orders this means that multiple flat fields must be taken in order to obtain good quality data in these regions. We found that although the colour of the lamp is relatively stable, the intensity of the lamp drops by approximately 10\% during the first 30 minutes after switching it on. We also use the Tungsten lamp along with a standard optical power meter (Thor Labs PM100D) to measure the spectrograph throughput from the bifurcated fibre input through the spectrograph to the detector of between 4 to 8\% depending on details of components.

\begin{figure}
\begin{center}
\includegraphics[width=75mm,angle=-0]{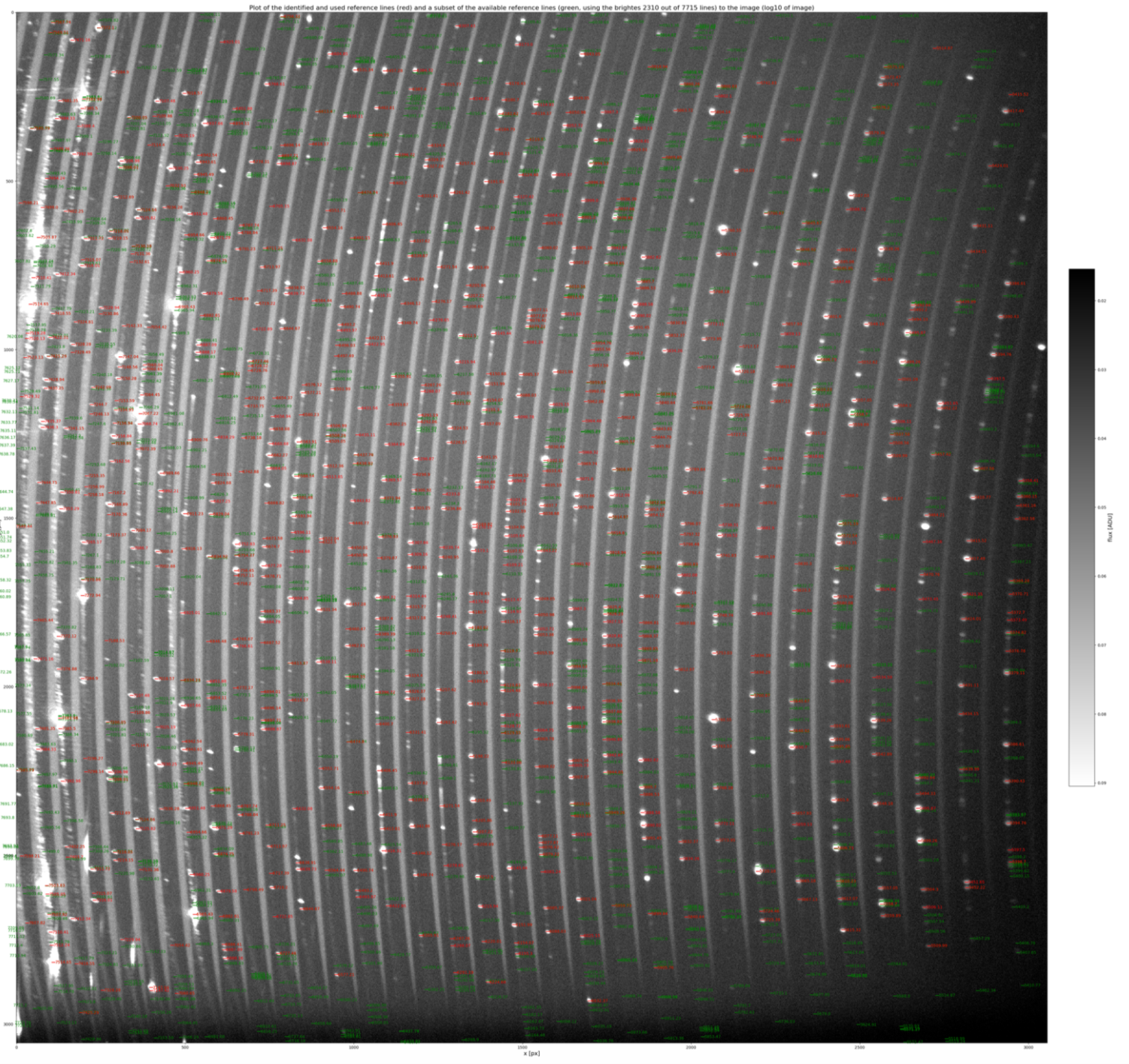}
\includegraphics[width=38mm,angle=-0]{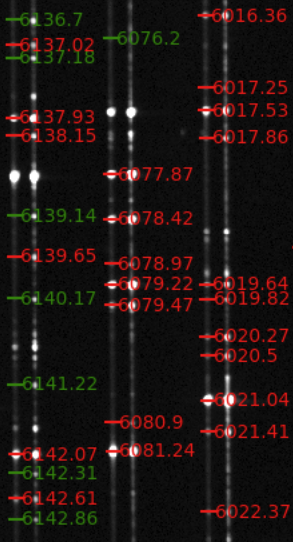}
\includegraphics[width=29.5mm,angle=-0]{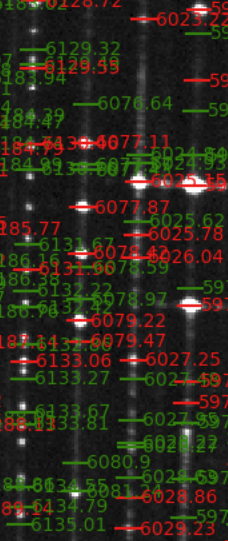}
\caption{{\bf The left-hand image shows an image (log10 gray scale) from our spectrograph showing the continuous orders of a flatfield lamp and simultaneously fed by a ThAr calibration lamp. The right-hand images show an image from HARPS (three orders with simultaneous ThAr in both fibres) on the left and our spectrograph image on the right (four orders with a single fibre input). ThAr lines are automatically identified by the pipeline} from the reference catalogue (in red) and used to create the wavelength solution. The remaining lines of the reference catalogue, not used for fitting the solution are shown in green.}
\label{lineid}
\end{center}
\end{figure}

For wavelength calibration reference we use a Thorium Argon lamp (P858A - hollow cathode lamp - Thorium with Argon gas fill from Photron). This typically gives around a few thousand potentially usable calibration lines across the optical regime, e.g., the left-hand side of Fig. $\ref{lineid}$ shows lines identified from the reference catalogue (in red) used to create the wavelength solution. The remaining lines of the reference catalogue, not used for fitting the solution, are shown in green. The right-hand side of Fig. $\ref{lineid}$ shows two images. The left-hand one is from HARPS (showing three orders with dual ThAr feed) and the right-hand one from this spectrograph (showing four orders). For the redder orders ThAr shows a smaller number of very bright lines and a comparison between ThAr and UrNe lines  (e.g., Ramsey et al. 2008) indicates  a greater line density would be available by using the larger line density available from a UrNe lamp and indeed the output from ThAr and UrNe; might easily be coupled together. However, it is likely that the development of laser combs and perhaps particularly more affordable devices like the Betters et al. (2016) photonic comb will be the most appropriate calibration.

\subsection{Fiber Input and Modal Noise}

We introduce light into the spectrograph by means of one fibre for the target and one for the reference source. This is achieved with two 10$\mu$m core 0.1 numerical aperture multimode fibres (Thorlabs FG010LDAFBUNDLE) supplied by Thorlabs as a ``custom 1$\times2$ fanout bundle'' from their range of bifurcated fibre bundles. Each fibre is contained in a protective jacket (Thorlabs FT061PS). These two fibres are combined in a common end connector jacket (also inside FT061PS protective cable) which is connectorised with  common end connectors (FC/PC, 30260G3). The reference fibre is connected directly into a ThAr or Tungsten flatfield lamp. The target fibre is usually connected into a Fiberbench Beamsplitter Module (Thorlabs FBT-50NIR- 50:50) which allows for flexible connection to sky fibre or another calibration source. We find that the fibres used (FG010LDA) have a peak throughput of 75\% where we measure the absolute throughput of fibres as part of our fibre characterisation experiments (e.g., Choochalerm et al. 2020)

We can get light from the Sun into the laboratory by imaging onto a 5~mm ball lens with a 10$^{\circ}$ field of view (Thorlabs 43-412) connected to a 15~m long 50~$\mu$m graded index 0.22 numerical aperture FC/PC cable (Thorlabs M42L15). The large field of view of the ball lens means that tracking errors should not compromise obtaining integrated solar disk light.  Unfortunately, we found that the field of view was so large that light scattered by nearby clouds noticeably degraded the quality of our signal and so observations were only attempted when skies were mostly clear.  Clouds that do not uniformly cover the Sun during an integration can lead to a significant solar rotation residual since the variation in the velocity across the disk of the Sun is 2~km/s. Thus, although the Sun is an obvious target for our experiments (e.g., Ramsey et al. 2008) this variation across the disk, along with clouds and a location which is rather prone to wind shake currently leads to an RMS between 10s exposures of 15 m/s but can only be maintained for a few minutes. We have achieved slightly better results on the Moon (14m/s with one minute exposures over 20 minutes). However, instabilities in our guiding and telescope equipment along with a lack of opportunity have so far limited our on-sky data.

\begin{figure}
\begin{center}
\includegraphics[width=120mm,angle=-0]{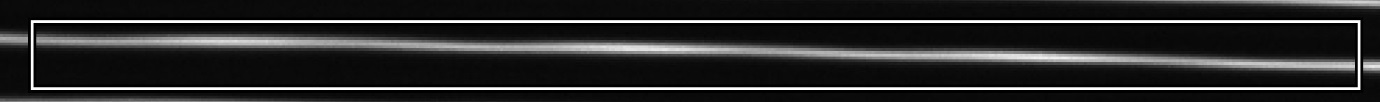}\\
\includegraphics[width=120mm,angle=-0]{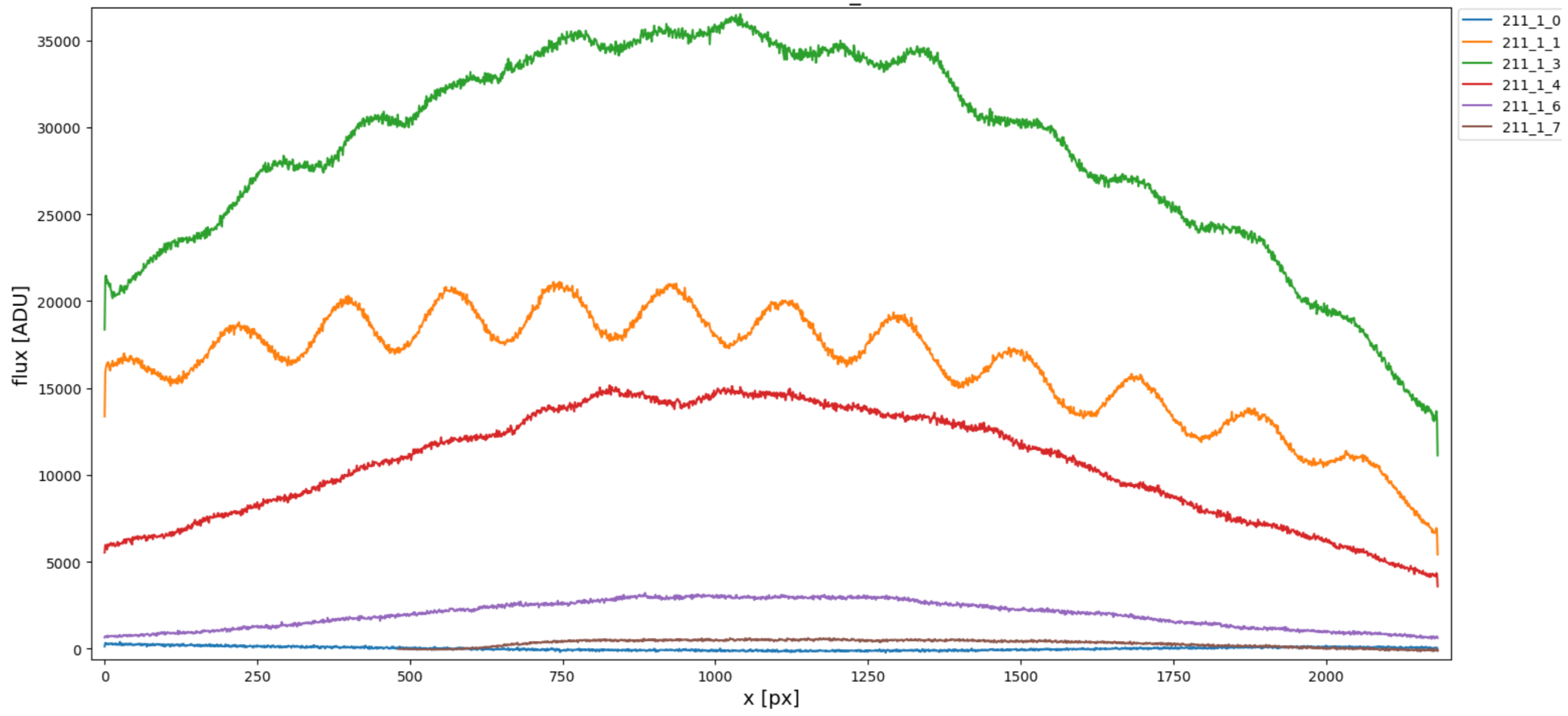}\\
\includegraphics[width=120mm,angle=-0]{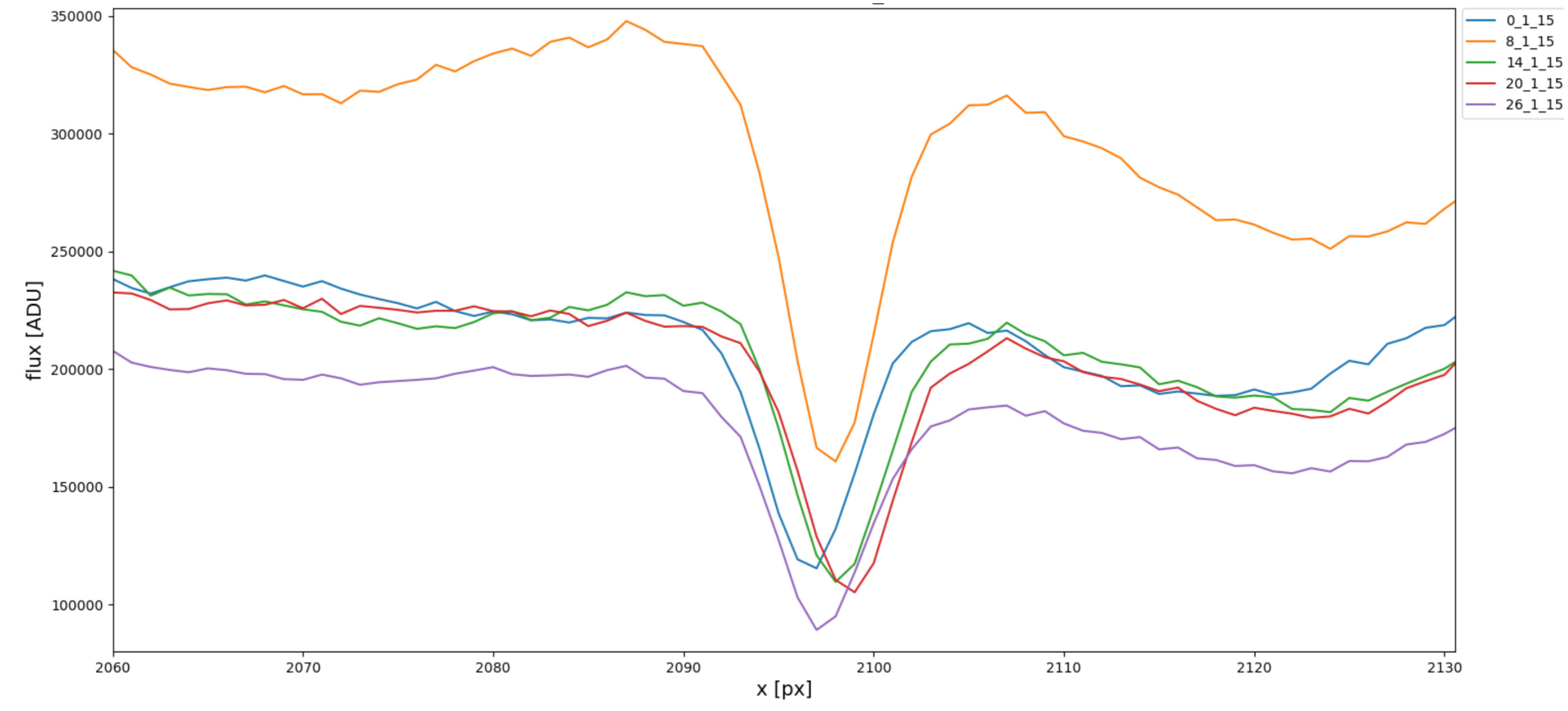}
\caption{The plots illustrate the impact of modal noise on the spectra. The upper plot shows a zoom in of a raw frame where the trace of spectrum (in white and grey against black background) can be seen as apparently periodic features of substantial strength when the spectra are scaled appropriately. These periodic features can also readily be been seen in the extracted spectra. The middle plot shows spectral extractions of different wavelength regions: (blue, 900~nm), 15 (orange, 760~nm), 30 (green, 650~nm), 45 (red, 570~nm), 60 (violet, 510~nm), 75 (brown, 460~nm). The lower plot shows a zoom in on an iron line (Fe${\rm\sc I}$ at 667.8nm) in the Sun prior to wavelength correction by ThAr where the fibre illumination has been adjusted to maximise the impact of modal noise on a spectral line. }
\label{modalnoise}
\end{center}
\end{figure}

We have found that in our instrumental setup, modal noise can be a significant issue. It can be a strong function of wavelength, temperature and even caused by the small agitation induced for example by the laboratory air conditioning fans. In the upper plot of Fig. $\ref{modalnoise}$ with appropriate scaling and zooming, modal noise can for some wavelengths even be seen in raw data from the camera. The upper plot of Fig. $\ref{modalnoise}$ presents an illustration of the wavelength dependence of the problem. The issue of modal noise is a well known one for high resolution spectrographs and has been examined in substantial detail; there are a number of potentially mitigations that may be made though no ``silver bullet'' (e.g., Ishizuka et al. 2016; Petersburg et al. 2018; Blinda, Conoda \& Wildi 2017). One of the best options would appear to be the appropriate agitation of the fibre in order minimize the effects of modal noise. While we did find this effective our available mechanical solution introduced unwanted vibrations into the setup. Thus instead we deploy an optical solution to the problem by using a mirror galvanometer system.

\begin{figure}
\begin{center}
\includegraphics[width=100mm,angle=-0]{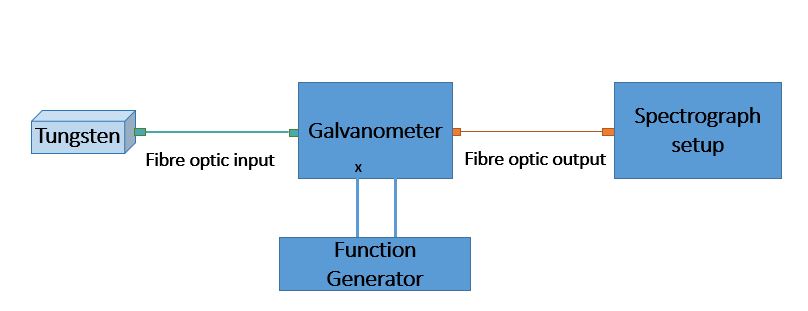}\\
\includegraphics[width=100mm,angle=-0]{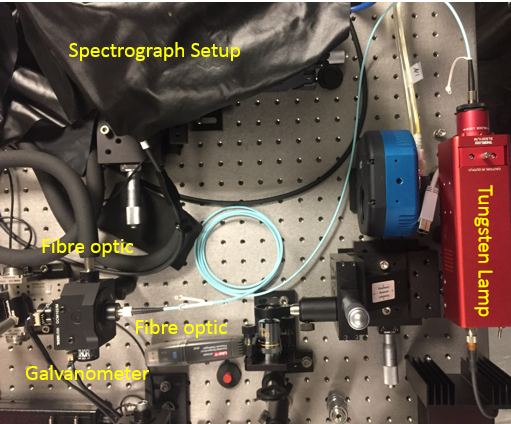}\\
\includegraphics[width=100mm,angle=-0]{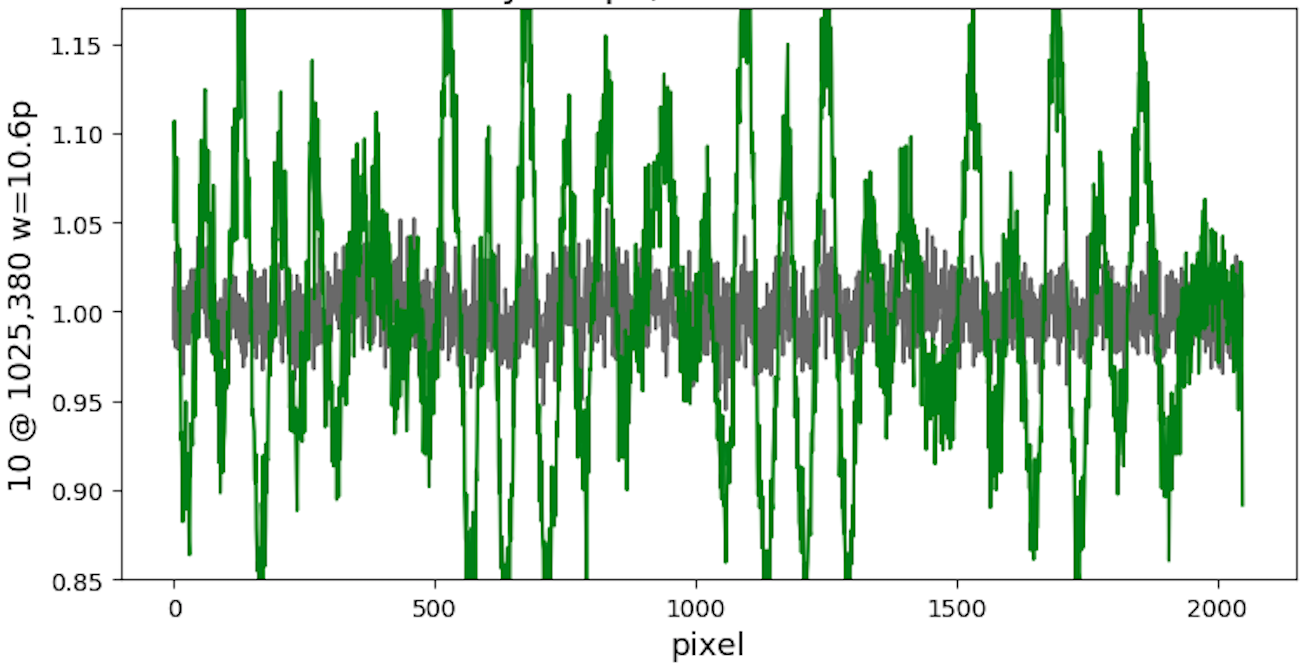}
\caption{The upper image shows the schematic layout of the spectrograph with galvanometer, the middle image shows the galvanometer on the bench (immediately above the written galvanometer label) and the lower plot gives an example of the impact of modal noise with and without the galvanometer operating.}
\label{galvo}
\end{center}
\end{figure}

We use silver coated mirrors (Thorlabs GVS002) and a function generator (Tektronix AFG1022) to apply appropriate patterns of frequencies and amplitude in order to apply a small movement to the beam so that the beam position is changed slightly and thus the illumination of the fibre carrying the light to the spectrograph is altered. We typically use this in 10-100~Hz range with voltages of 20--50~mV. In Fig. $\ref{galvo}$, the top plot shows a schematic of the arrangement used to test the impact of modal noise, the middle plot shows the reality on the optical bench of the different components and the lower plot illustrates the improvement available by modulation of the beam. While this system is effective, we note that in order to optimise the reduction in modal noise, we need to introduce very significant movement of the beam which then causes a significant loss of throughput. In the case of this plot by a factor of about two in throughput and so this system requires further optimisation (Boonsri et al. 2020, in preparation) and may not produce as robust results as might be achieved with a double scrambler. Many of our problems are known consequences of using such small fibres. For example they are more difficult to handle and more prone to focal ratio degradation and guiding errors (e.g., Kannappan, Fabricant \& Hughes 2002). They also provide significant differences in potential signal-to-noise as a function of core diameter and geometry (e.g., Sablowski et al. 2016). These issues are closely related to our fibre tapering experiments (Choochalerm et al. 2020).

\subsection{Active Stablisation}

The beam is turned through 90$\deg$ toward the camera using a right-angle prism mirror (Thorlabs MRA20-G01). It is mounted on an XY translation stage consisting of a Kinematic Mount (KM200T - SM2), a Kinematic Prism Mount (KM200PM/M) and a Reflecting mirror Mount (KM200CP/M). Its position can be adjusted by piezo inertial actuators (Thorlabs PIAK10) which offer movements of 20 nm and ``no backlash'' in a compact package. 

\begin{figure}
\begin{center}
\includegraphics[width=160mm,angle=-0]{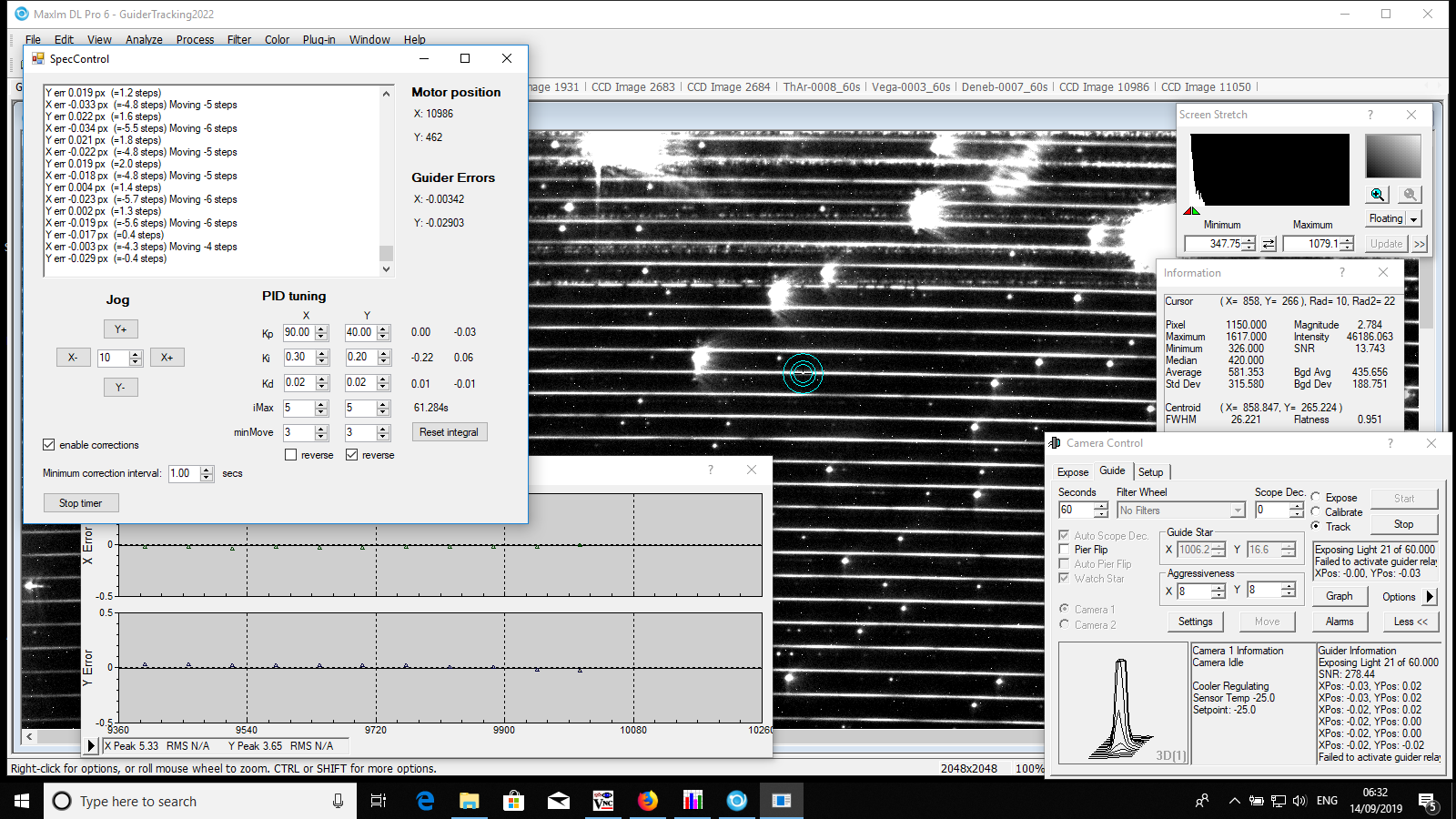}
\caption{Screen capture of the control computer showing the feedback system in operation. The main frame shows the raw image frame with a flat field lamp (a Thorlabs OSL1-EC with low red throughput enabling arc lines to be easily seen) and ThAr both in the target fibre. The camera control window in the top right indicates among other things that the image (an arc line) is being tracked (as a guide star) and indicates the camera information along with centroid displacement of the image. The lower left window indicates the corrections made by the piezo actuators based on the deviations of the centroid reported in the camera control window. The bottom right plot shows two separate plots of the displacement of the arcline in $x$ and $y$ with time. Once tracking has been switched on a fairly rapid adjustment of the centroid position occurs to bring it to (0,0). }
\label{closedloop}
\end{center}
\end{figure}

Our active stabilisation system uses telescope guiding software (MaxIm DL\footnote{\url{https://diffractionlimited.com/product/maxim-dl/}}) to detect the position of a ThAr line and this ``star like'' reference image is used as a ``guide star''.  The ThAr image position on the detector is then used to control XY actuators (Thorlabs) on the turning mirror to stabilise the spectrum of the object on the detector.  Fig. $\ref{closedloop}$ shows a screen capture of the control computer, the camera control window in the top right indicates, among other things, that the image (an arc line) is being tracked and it displays the camera information along with centroid displacement of the image. The other windows monitor the image. In the top left is the image that is being tracked along with a circular apertures for which statistics are reported in the top middle window. The lower right window indicates the corrections made by the actuator based on the errors of the centroid reported in the camera control window. The bottom left plot shows the centroid behaviour in $x$ and $y$ with no tracking and then once tracking has been triggered, a fairly rapid adjustment of the centroid position to the chosen baseline can be seen.

The guiding loop to control the actuators is written in visual basic using the .Net APIs and uses the Visual Studio along with the guiding function of MaxIm DL. This can be used in single or multi-guide star mode.  MaxIm DL allows a variety of different guiding functionalities of guiding from automated single- or multi-star to fully defined by the user. MaxIm DL is used in ``single-'' or ``multi-star'' autoguiding mode to capture the images and measure the positions of the "guide stars". The guiding loop to control the actuators is written in C$\#$ and is available on Github\footnote{\url{https://github.com/bayfordbury-observatory/EXOhSPEC}}. It uses the Thorlabs Kinesis .Net APIs and uses MaxIm DL's COM interface to access the guiding errors as the input to a PID control loop. The errors are dependent on the exposure (signal to noise) so optimisation of time constants is required. Our current implementation of the loop uses K$\rm{p}$ proportional terms (x=10, y=20), K$\rm{p}$ integral (steady state error) terms ($x$=0.2, $y$=0.2), K$\rm{p}$ derivative (dampening) terms ($x$=0.1, $y$=0.1), an iMax limit for integral to prevent wind-up of $x+y$ = 1, and a Minimum Move of $x+y$ = 2.  The control loop code is provided at 
 
\begin{figure}
\begin{center}
\includegraphics[width=180mm,angle=-0]{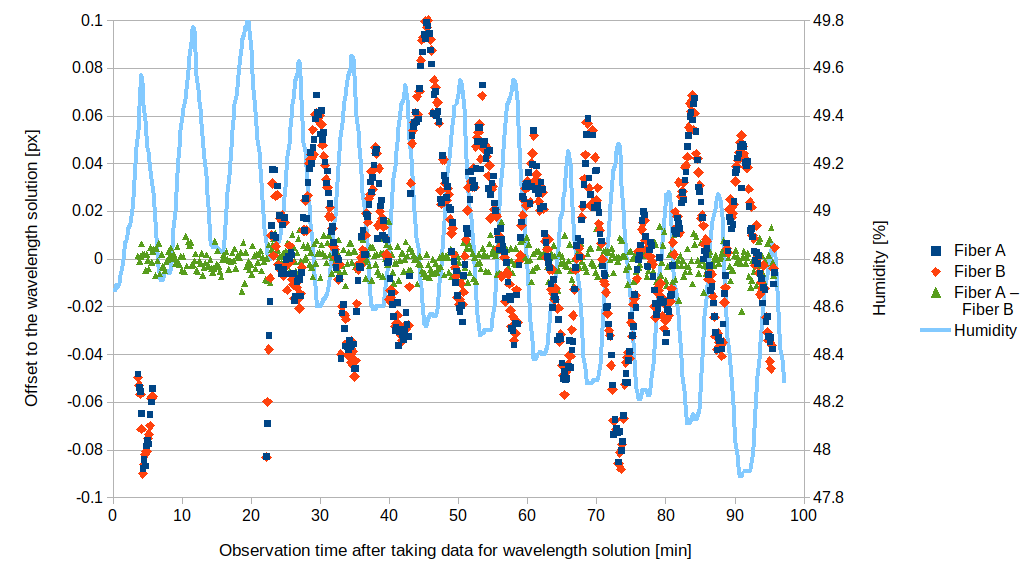}\\
\caption{Offsets to the wavelength solution caused by humidity changes in the spectrograph arising from the laboratory air conditioning system over a timescale of 100 minutes when both fibres are fed by the same calibration lamp. }
\label{aspiration}
\end{center}
\end{figure}

One of the key facets of a radial velocity spectrograph is long-term temperature, pressure and humidity stability and to this end modern spectrographs typically work in a vacuum and are specified to have mK temperature control. In order to check the stability of the current setup, we split the light from the ThAr lamp into two beams and feed the target and calibration fibres separately with the same lamp. This procedure is similar to the nightly calibration procedure of HARPS\footnote{\url{https://www.eso.org/sci/facilities/lasilla/instruments/harps/inst/calibrations.html}}. We used continuous 10~second ThAr exposures. Fig. $\ref{aspiration}$ shows the measured difference in radial velocity between the target and calibration fibres over 100 minutes. The control loop was switched off until 22 minutes, hence creating offsets of several tens of a pixel (beyond the plotted scale). Once switched on the control loop effectively corrects for the changes induced by the regular cycle of the air conditioning system, as measured by the change in humidity. The green points show difference in signal between the two calibration fibres. The RMS of the green points over the period is 0.005 pixel. However, with the current setup small variations can't be corrected due to the response time of the loop and hysteresis of the elements. When the air conditioning system is switched off the RMS can be reduced further, however, an RMS below 0.0035 pixel (4~\ms) is not reliably achieved with the current set up. Optically, the axis alignment of the actuators is currently approximately 60 deg from one another, rather than orthogonal. Also the detailed performance of these actuators is not linear at small displacements due to friction and hysteresis. The current step size of the actuators is also rather coarse as one step is approximately 0.005 pixels. Although this and the current control loop settings work relatively well to follow the humidity changes, they can over- and under-correct and the system needs to be redesigned to reach a significantly higher level of sensitivity. 

We can compare these results with those found for HARPS calibration. Following a similar analysis using routine calibration data taken from the HARPS archive, we find the same procedure yields an RMS scatter of 0.7~milli-pixel (0.6~\ms). For this experiment, the beam splitter had a rather low throughput in the blue wavelength range and so only lines redder than 510~nm were available for the wavelength solution, which also adds error to the quality of the underlying wavelength solutions which are being compared. For example, looking at the same wavelength range and using the same HiFLEX pipeline (Errmann et al. 2020), the wavelength solution for the EXOhSPEC prototype finds 540 lines to use, whereas for the HARPS data finds 1930 lines to use. The difference in the number of lines can be attributed to a number of factors perhaps particularly the optics (especially the use of doublet lens), quality of the ThAr lamp (Lo Curto 2017) but more simply our short ThAr exposures. For these experiments the sCMOS camera used a firmware version with a maximum of 10s exposure rather than 40s used by HARPS. Longer exposures would enable the robust identification of many fainter lines which become available for longer ThAr exposures. 

For our current implementation of the prototype we rely on MaxIm DL software which brings some compromises. For example, the range of cameras for which software drivers exist is limited. In the terms of the active stabilisation system we have tried the available default guiding options and find improvement in consistency from using ``multi-star'' (multiple ThAr lines) rather than the other available default options. However, more robust guiding and thus better stability might also be available by optimisation of the corrective image shifts given to the actuators by Visual Studio. For example, image processing outside of MaxIm DL might follow a methodology like (Gunther et al. 2017) and limit the lines used for centroiding to an optimised set of ThAr lines (e.g., based on Lovis \& Pepe 2007).

\section{Conclusion}

The process of building a small stable high resolution spectrograph using commercially available components has involved moving from (1) a variety of SLR camera lenses to an achromatic doublet, (2) standard input fibre, to a birfurcated fibre, (3) iteration of prisms for improved anti-reflection coating and appropriate cross-dispersion properties, (4) a replica grating from Thorlabs to larger more efficient Richardson one, (5) control of modal noise through introduction of a galvanometer, (6)  use of a CMOS rather than a CCD camera and (7) inactive to active control with second order correction. We have explored a number of system options and demonstrate how a relatively small and inexpensive spectrograph built from off-the-shelf components can achieve high resolution and reasonable optical quality as judged relative to the HARPS spectrograph. A key finding is that precisions below 5\ms~can be achieved by using commercial autoguiding software to determine the appropriate feedback to maintain stability. Further work is required to demonstrate robust system performance in an astronomical setting on a telescope.

\section{Acknowledgements}
We are grateful for support of this project through a Newton fund grant from the UK Science Technology and Research Council (ST/P005667/1 and ST/R006598/1), to John Collins for his donation of his Celestron Edge HD, to Peter Beck for his donation of the Tungsten lamp to the project and to Christophe Buisset and Suparerk Aukkaravittayapun for supporting this project at NARIT. GAE acknowledges a Ramón y Cajal Research fellowship. We are glad to acknowledge the use of calibration data obtained from the ESO Science Archive Facility from proposal 072.C-0488(E) based on nights 2004-07-10 to 2004-07-31. The original version of this manuscript was significantly improved by the insights of the anonymous referee.

\section{References}
\noindent
Baker, C., Anglada-Escude, G., Jones, H., Martin, W.E., 2019, SPIE, 11117, 111171 \\
Becker, J.M., 1976, Nature, 260, 227\\
Betters C. H., Hermouet M., Blanc T., Colless J. I., Bland-Hawthorn J., Kos J., Leon-Saval S., 2016, SPIE, 9912, 99126 \\
Blinda, N., Conoda, U., Wildi, F., 2017, AO4ELT5 Conference series, (arXiv:1711.00835)\\
Butler, R. P., Marcy, G. W., Williams, E., McCarthy, C., Dosanjh, P., Vogt, S. S., 1996, PASP, 108, 500\\
Butler, R.P., Jones, H.R.A., Feng F. et al. 2019, AJ, 158, 6 \\
Bond, C, Brown, D., Freise, A., Strain, K.A., 2016, Living Reviews in Relativity, 19, 3 \\
Buikema, A. et al., 2020, arXiv:2008.01301 \\
Calderone G. et al. 2016,  Software and Cyberinfrastructure for Astronomy IV, SPIE, 99132\\
Campbell B., Walker~G.A.H., 1979, PASP, 91, 540\\
Campbell B., Walker G.A.H., Yang S., 1988, ApJ, 331, 902\\
Choochalerm P., et al., 2020, Optical Fiber Technology, submitted \\
Cosentino, R., Lovis, C., Pepe, F., 2012, SPIE, 8446, 84461V \\
Crane, J.D., S. A. Shectman, S.A., Butler, R.P., 2006, SPIE 6269, 626931\\
Errmann R., et al., 2020, PASP, 132, 064504 \\
ESA, 2014, JUICE Study Definition Report, ESA/SRE\\
Feng, F., et al., 2019, ApJS, 244, 39 \\
Gunther, M. et al., 2017, MNRAS, 472, 295\\
Haynes, D., Haynes, R., Olaya J.C., Laval, S., 2012, SPIE, 8450, 84503\\
Ishizuka M., Kotani T., Nishikawa J., Tamura M., Kurokawa T., Mori T., Kokubo T., 2016, SPIE, 99121 \\
Kannappan, S.J., Fabricant D.G., Hughes C.B., 2002, PASP, 114, 795\\
Kawinkij A. et al., 2019, SPIE, 11116, 111161G\\
Koch~A., Woehl~H., 1984, A\&A, 134, 134\\
Latham, D.W., et al., 1989, Nature, 339, 38\\
Lhospice, E. et al., 2019, SPIE, 11117, 111170Z \\
Lizon, J.L. et al., 2016, SPIE,  9908, 990866 \\
Lo Curto, G., 2017, Instrument talk : HARPS. Zenodo. (http://doi.org/10.5281/zenodo.887265)\\
Lovis, C. \& Pepe, F., 2007, A\&A, 468, 1115\\
Lovis, C. et al., 2017, A\&A, 599, 16\\
Pepe, F., et al., 2000, SPIE, 4008, 582\\
Pepe, F. et al., 2020, arxiv:2010.00316 \\
Peters, A., Chung K.Y., Chu, S., 2001, Metrologia, 38,25 \\
Petersburg R.R., McCracken T.M., Eggermann D., Jurgenson C..E., Sawyer D., Szymkowiak A.E., Fischer D.A., 2018, ApJ, 853, 181\\
Ramsey, L. et al., 2008, PASP, 120, 887\\
Sablowski, D.P., Pluschke, D., Weber, M., Strassmeier K.G., Jarvinen, A. , 2016, AN, 337, 216\\
Schroeder D.J., 1987, Asronomical Optics, Academic Press (San Diego)\\
Schroeder, D.J., 1967, ApOpt, 6, 11\\
Schroeder, D.J., 1970, PASP, 82, 1253 \\
Schroeder, D.J., Anderson, C.M., 1971, PASP, 83, 43\\
Seifahrt, A. Stürmer, J., Bean, J.L., Schwab, C., 2018, SPIE, 10702, 107026 \\
Struve O., 1952, Observatory, 72, 199 \\
Tull, R.G. 1998, SPIE, 3355, 387\\
Vogt, S.S. et al, 1994, SPIE,  2198, 362\\
Wolszczan, A., Frail, D.A., 1992, Nature, 355, 145\\
Zhan, H., Zhang, X. Cao, L., 2017, JINST, 12, C04010 \\

\end{document}